\documentclass[twocolumn,showpacs,amsmath,amssymb,aps,prd]{revtex4}

\usepackage{bm}

\begin{document}

\title{What is the maximum rate at which entropy of a string can increase?}

\author{Kostyantyn Ropotenko}
 \email{ro@stc.gov.ua}

\affiliation{State Administration of Communications, Ministry of
Transport and Communications of Ukraine 22, Khreschatyk, 01001,
Kyiv, Ukraine}

\date{\today}

\begin{abstract}
According to Susskind, a string falling toward a black hole spreads
exponentially over the stretched horizon due to repulsive
interactions of the string bits. In this paper such a string is
modeled as a self-avoiding walk and the string entropy is found. It
is shown that the rate at which information/entropy contained in the
string spreads is the maximum rate allowed by quantum theory. The
maximum rate at which the black hole entropy can increase when a
string falls into a black hole is also discussed.
\end{abstract}


\maketitle

The paradox of information loss in black hole physics \cite{s1} is
essentially related to the flow of information or, taking into
account the fundamental equivalence relation between information and
entropy, information and entropy. Its resolution requires a
kinematic description of matter at high energies which differs
radically from the one offered by conventional quantum field theory.
String theory is widely believed to provide such a description. In a
series of insightful papers, Susskind showed \cite{s2,s3} that a
string falling toward a black hole spreads linearly over the
stretched horizon. But during the process string interactions become
important so that the process turns out to be nonperturbative
\cite{s1}. A calculation of such processes is beyond the current
technology of string theory. Susskind suggested that in this case a
string should spread much more rapidly - exponentially \cite{s3}.
Obviously that during the process the entropy of string should
change. But Susskind did not consider the string entropy and present
a calculation of its change. In this paper I find the entropy of
spreading string with the help of random-walk models. As is well
known, quantum theory, and, in particular, quantum information
theory impose constraints on the information/entropy flow. String
theory pretends to be a consistent quantum theory of matter. But it
is unclear whether string theory satisfies quantum limits to the
entropy flow when a string spreads exponentially. To answer the
question I find the maximum rate allowed by quantum theory at which
the entropy of a string can increase when it spreads over the
stretched horizon of a Schwarzschild black hole.

Before we start out discussing the problem, I repeat, for the
convenience of the reader, some well-known facts from \cite{s1}
concerning the behavior of stringy matter near the horizon without
proofs, thus making our exposition self-contained. It is well
established that, from the point of view of an external observer,
the classical physics of a quasistationary black hole can be
described in terms of a "stretched horizon" which is a "membrane" -
a timelike surface placed near the event horizon and endowed with
certain mechanical, electrical, and thermal properties \cite{tor}.
The exact distance of this membrane above the event horizon is
somewhat arbitrary. In the context of string theory - the subject of
our research, the stretched horizon is most naturally thought of as
lying at the string scale $l_s$ above the event horizon. In what
follows we will deal with this conception (sometimes referred to
simply as horizon). An important fact is that the size and shape of
a string are sensitive to the time resolution. It is a smearing time
over which the internal motions of the string are averaged. Susskind
showed \cite{s2} that zero-point fluctuations of a string make the
size of the string depend on a time resolution; the shorter the time
over which the oscillations of a string are averaged the larger its
spatial extent. This effect is closely related to the well-known
Regge behavior of string scattering amplitudes \cite{s2}. More
precisely, Susskind found that in the weak coupling limit the mean
squared radii of the string in the transverse and longitudinal
directions in Planck units are $\langle R_{\perp}^{2}\rangle
=l_s^{2}\ln (1/ \tau _{res})$ and $\langle R_{\parallel}^{2}\rangle
=l_s^{2}/\tau _{res}$, respectively. Similar calculations can be
performed for the total length of the string, and Susskind found
that $L=l_s /\tau _{res}$. Now consider a string falling toward a
black hole. As is well known \cite{s1}, the proper time in the frame
of the string $\tau$ and the Schwarzschild time of an external
observer $t$ are related through $\tau \sim \exp({-t/2R_{g}})$  due
to the redshift factor. This means that the transverse size of the
string will increase linearly:
\begin{equation}
\label{str5}\langle R_{\perp}^{2}\rangle =l_s^{2}\frac{t}{2R_{g}},
\end{equation}
while its longitudinal size and total length - exponentially:
\begin{equation}
\label{st1}\langle R_{\parallel}^{2}\rangle
=l_s^{2}\exp({t/2R_{g}}),
\end{equation}
\begin{equation}
\label{st2}L =l_s\exp({t/2R_{g}}).
\end{equation}
But the longitudinal growth (\ref{st1}) is rapidly canceled by the
Lorentz longitudinal contraction. Thus the string approaching the
horizon spreads only in the transverse directions (in this
connection the subscript "$\perp$" at the the mean squared radius of
the string will be replaced by the "s", that means "string",
henceforward).  Note that as Mezhlumian, Peet and Thorlacius showed
\cite{mez}, the transverse spreading can be also described as a
branching diffusion of wee string bits.

\emph{How does the entropy of string change}?- Unfortunately
Susskind did not consider the string entropy and give a calculation
of its change. To answer the question I propose to use the
random-walk model. As has been stated above, Susskind obtained his
results in the framework of weakly coupled string theory. It turns
out that behavior of a string in this regime is very precisely
described in terms of the random-walk model \cite{zwie},
\cite{thor}. So we can imagine a string as simply built by joining
together bits of string, each of which is of length $l_s$. Suppose
that the total length of the string is $L$ and each string bit can
point in any of $n$ possible directions. Then the number of bits is
$N=L/l_s$ and the number of states of the string is
\begin{equation}
\label{str1} c_N \sim n^{N}\sim \exp {(L\ln n/l_s}).
\end{equation}
For notational simplicity the factor $\ln n $ will be omitted
henceforward (there is no loss of generality in doing so because we
can always redefine $g$, $l_S$ and $L$). We can also define the mean
squared radius of the string $\langle R_s^{2}\rangle = Ll_s$. Then
\begin{equation}
\label{str2} c_N  \sim \exp {(\langle R_s^{2}\rangle/l_s^{2})},
\end{equation}
and the entropy of the string is
\begin{equation}
\label{str3} S_s \sim L/l_s \sim \langle R_s^{2}\rangle/l_s^{2}.
\end{equation}
Now, substituting (\ref{str5}) in (\ref{str2}), we obtain
\begin{equation}
\label{str6} c_N  \sim \exp {(\langle R_s^{2}\rangle/l_s^{2})}\sim
\exp(t/2R_{g}).
\end{equation}
So the entropy is
\begin{equation}
\label{enstr} S_s \sim \frac{t}{2R_{g}},
\end{equation}
and the entropy rate
\begin{equation}
\label{rate1} \frac{dS_s}{dt}=\frac{1}{2R_{g}}.
\end{equation}
Thus during the spreading the entropy of string increases. It is
obvious from the thermodynamical point of view: the spreading effect
is a result of heat exchange between a black hole and a string.
Since the temperature of the black hole radiation depends on the
radial position, $T(r)=T_H/\chi$, where $T_H$ is the Hawking
temperature, $T_H=1/4 \pi R_g$, and $\chi$ is the the redshift
factor, $\chi=(1-R_g/r)^{1/2}$, it follows that from the viewpoint
of the external observer the string falls into an increasingly hot
region. So there is a flow of heat from the black hole to the
string. Thus the string should "melt" and spread throughout the
horizon. Obviously during this process the phase volume and entropy
of string increase. Susskind demonstrated the spreading effect for a
fundamental string. It is widely believed, however, that it is not a
peculiar feature of a special (still hypothetical) kind of matter.
Susskind suggested that in the framework of the so-called
infrared/ultraviolet connection \cite{s1} it is a general property
of all matter at energies above the Planck scale.

From (\ref{str6}) we have found the entropy of string in terms of
the mean squared radius or area. According to (\ref{str1}) we can
also do it in terms of the total length. But in this case
\begin{equation}
\label{ad1} c_N \sim \exp(\exp (t/2R_{g})),
\end{equation}
and for the string entropy we obtain a different result:
\begin{equation}
\label{entrop} S_s \sim \exp ({t/2R_{g}}).
\end{equation}
So the relations (\ref{str3}) are not satisfied. The point is, as
noted by Susskind himself \cite{s3}, that the linear growth of area
(\ref{str5}) was obtained in the framework of free string theory. It
does not take into account such a nonperturbative phenomenon as
string interactions. The exponential growth of string length and
linear growth of area imply that the transverse density of string
should increase. When the density becomes of order $1/G$, the area
density of horizon entropy, interactions become important. But the
precise calculation of the spreading effect in the nonperturbative
regime is beyond the current technology of string theory. To prevent
the density from increasing beyond $1/G$, Susskind suggested
\cite{s3} that the nonperturbative effects must be such that the
string bits become repulsive. This will produce an outward pressure
that spreads the string bits much more rapidly than the linear
growth in the free theory; as a result, a true growth must be
exponential:
\begin{equation}
\label{exp}\langle R_s ^{2}\rangle =l_s^{2}\exp({t/2R_{g}})
\end{equation}
(at the same time the total length retains its form (\ref{st2})). It
has been known for a long time that a classical charged particle
spreads exponentially on the stretched horizon \cite{tor}. So it is
reasonable that these growth patterns (classical and stringy) are a
match. That is, (\ref{exp}) is the only reasonable assumption which
provides this match. Now the entropy expressed in terms of the
length and area has the same form and the relations (\ref{str3}) are
satisfied. Thus for the entropy rate we obtain
\begin{equation}
\label{rate2} \frac{dS_s}{dt}= \frac{S_s}{2R_{g}}.
\end{equation}
It seems however that this is a comparatively large rate. Perhaps
the point is that we have used the formulas of the simple
random-walk model which is valid only in the framework of the free
string theory? So we are forced to use more suitable models. Since
the string bits become repulsive, this effect can be accounted for
by imposing the condition that two bits cannot occupy the same site.
In polymer physics this type of condition is called the "excluded
volume effect" \cite{gen}. If we model a string as a connected path
on the stretched horizon, the excluded volume effect will be
correspondent to the condition that the path cannot pass through any
sites that have been traversed previously. In mathematics this is
called a "self-avoiding walk" \cite{walk}. So our simple model
should be replaced by the model of self-avoiding walks. At first
sight it seems that the repulsive interactions will impose
constraints on the total number of string states and the huge
entropy rate (\ref{rate2}) should considerably reduce. It turns out
that this is not the case. The mathematical properties of simple
random walks are trivial, but the mathematical properties of
self-avoiding walks are complex. The conjectured asymptotic number
of the self-avoiding walks of $N$ steps is \cite{gen,walk}:
\begin{equation}
\label{num} c_N=\mbox{constant}\,\, \tilde n^{N}N^{\gamma -1}.
\end{equation}
The first factor $\tilde n^{N}$ is reminiscent of the $n^{N}$ in
(\ref{str1}), but $\tilde n$ is somewhat smaller than $n$; the exact
value of $\tilde n$ is not known for the hypercubic  lattice in any
dimensions $\geq 2$, although for the honeycomb lattice in two
dimensions there is nonrigorous evidence that $\tilde
n=\sqrt{2+\sqrt{2}}$. The exponent $\gamma \cong 4/3$. So, despite
string interactions, the leading significant term of the total
number of allowed string states retains its previous form
(\ref{str6}). Thus the entropy rate is really (\ref{rate2}). String
theory pretends to be a consistent quantum theory of matter. But we
do not know whether the rate (\ref{rate2}) satisfies quantum limits
to the entropy flow.

\emph{Quantum limits to the entropy rate.}- As mentioned in the
beginning quantum information theory imposes quantum limits to the
information/entropy flow. The Bekenstein-Bremermann limit
\cite{bre}, \cite{bek} is one of the most important limits of just
the same kind. In its original form it sets an upper bound on the
rate at which information $I$ may be transmitted by a signal with a
certain amount of energy $\Delta E$,
\begin{equation}
\label{inf1} \dot{I}_{max} \leq \frac{2\pi\Delta E}{\hbar}\log_{2}e
\hspace{0.5cm}\mbox {bits $s^{-1}$}
\end{equation}
(where the dot, as is customary in information theory, denotes
differentiation with respect to $t$ and the Planck constant is
written explicitly to emphasize the quantum nature of the bound).
Now I want to use it to check the conclusion of string theory
(\ref{rate2}). For this purpose let us rewrite the bound
(\ref{inf1}) in terms of entropy. According to the fundamental
equivalence relation between information and entropy, (\ref{inf1})
can be viewed as an upper bound on the rate of entropy flow in a
system
\begin{equation}
\label{inf2} \dot{S}_{max} \leq \frac{2\pi\Delta E}{\hbar}.
\end{equation}
Taking into account the second law of thermodynamics we obtain
\begin{equation}
\label{inf3} \dot{S}_{max} \leq \frac{2\pi T \Delta S}{\hbar},
\end{equation}
where $T$ and $\Delta S$ are the system's temperature and entropy
change, respectively. In this form it can be immediately applied to
a spreading string. We should, however, emphasize here the
following. The spreading process begins to occur when the string
reaches the horizon at a distance of order of the string scale
$l_{s}$ from the horizon in a thin layer $\sim l_{s}$. So it may
seem that the system's temperature should be determined by the
Hagedorn temperature, $T_{Hagedorn}= 1/2\pi l_s$. But for an
external observer it becomes the Hawking temperature due to the
gravitational redshift: since $\chi=(1-R_g/r)^{1/2}\approx l_s/2R_g$
at the proper distance $l_s$ from the horizon, $T_H=\chi
T_{Hagedorn}$. The entropy is not redshifted because it is an
invariant. So for the exponential spreading (\ref{entrop}) we get
\begin{equation}
\label{inf3} \dot{S}_{max} \leq \frac{2\pi S_sT_H}{\hbar}.
\end{equation}
Comparing it with (\ref{rate2}) expressed in terms of the Hawking
temperature with the explicitly written Planck constant
$T_H=\hbar/4\pi R_g$, we conclude that the rate (\ref{rate2})
satisfies the bound (\ref{inf3}). More precisely the rate
(\ref{rate2}) saturates the bound (\ref{inf3}). Thus (\ref{rate2})
is the maximum rate allowed by quantum theory at which the entropy
of a string can spread over the horizon.

Before now we have been ignoring the concept of a communication
channel: the Bekenstein-Bremermann limit does not use this concept
explicitly. But it is one of the most important concepts of
information theory. A communication channel is understood to be a
complete set of unidirectionally propagating modes of some fields,
with the modes enumerated by a single parameter \cite{pen}. Pendry
proposed a limit to information/entropy flow which is essentially
based on the concept of a communication channel \cite{pen}.
According to Pendry \cite{pen}, the maximum rate at which
information and entropy can flow in a channel is only determined by
the system's temperature $T$:
\begin{equation}
\label{pen1} \dot{I}_{max} \leq \frac{\pi T}{3\hbar}\log_2 e \quad
\mbox{or}\quad \dot{S}_{max} \leq\frac{\pi T}{3\hbar}
\end{equation}
(for a fermion channel it is reduced by a factor $\sqrt{2}$). It can
be immediately applied to the spreading process. But it is written
for one channel. To apply Pendry's limit to the spreading process we
should determine the real number of channels for a string. As
mentioned above, the number of channels is determined by the number
of modes; moreover each of the possible polarization states can be
used as a separate channel. Pendry showed \cite{pen} that
communication systems operate on a finite number of easily
identified channels to each of which the bound (\ref{pen1})
separately applies. For example, consider a flow of entropy
conducted by phonons in an insulating crystal \cite{pen}. In this
case the number of channels is given by $3n_AN_c$, where $n_A$ is
the number of atoms per unit cell and $N_c$ is the number of unit
two-dimensional cells of the crystal lattice projected onto the
plane perpendicular to the direction of flow. In our model a string
with a length $L$ has $3(2L/l_s)$ modes (taking account three
possible polarizations). By additivity the total entropy rate equals
that of one channel (\ref{pen1}) multiplied by the number of
channels. Then taking into account (\ref{str3}) we obtain
\begin{equation}
\label{rate4} \dot{S}_{max} \leq \frac{2\pi T \Delta S}{\hbar}.
\end{equation}
This is a quantum bound on the entropy flow expressed in terms of
the  channel. It has the same form as the bound (\ref{inf3})
obtained with the help of the Bekenstein-Bremermann limit. Obviously
the rate (\ref{rate2}) complies the bound (\ref{rate4}). Note that
as Pendry showed \cite{pen}, present communication systems for
transporting information are very far from the quantum limit
(\ref{rate4}). This is also true for the entropy flow in any
physical process in a laboratory. By contrast, the rate
(\ref{rate2}) saturates the bound (\ref{rate4}).

\emph{The maximum rate at which the black hole entropy can
increase}.- Since the area of the horizon $A$ is finite, it follows
from (\ref{exp}) that a string spreads over the entire horizon in a
finite time $t_{spread}= 2R_{g}\ln (A/l_P^{2})$ \cite{s3}. At this
time  the number of states of the string becomes $c_N\sim
\exp(R_g^{2}/l_P^{2})$ and the entropy of the string reaches that of
the black hole,
\begin{equation}
\label{BH} S_s = S_{bh}=\frac{A}{4l_P^{2}}.
\end{equation}
According to Susskind \cite{s3}, at the time $t_{spread}$ the string
completely covers the entire horizon and can no longer expand due to
the nonperturbative effects. The spreading ends and only a new
falling string or any other perturbation can start it again. The
next string falling toward the horizon interacts with a previous one
lying on the horizon in such a way that the formation of a single
(new) string is thermodynamically favored, etc \cite{s1}. So the
stretched horizon is a single string made out of all strings
whenever fallen into it. From the string theory point of view, a
black hole is nothing but a single string lying on the sphere of the
radius $R_g$. Since the spreading effect takes place on a short time
scale compared to the black hole lifetime $\sim R_{g}^{3}$, Susskind
\cite{s3} restricted himself to a static metric. But since a black
hole absorbs a string, its gravitational radius must increase.
According to the teleological nature of the event horizon
\cite{tor}, before a fall of the next string, the gravitational
radius and the horizon area increase like $\exp {(t/2R_{g})}$. The
same exponential growth governs the spreading of a classical
perturbation on the stretched horizon. This means that the spreading
effect takes place. As a result the black hole entropy increases
with the rate
\begin{equation}
\label{rate5} \frac{dS_{bh}}{dt} = 2\pi S_{bh}T_H.
\end{equation}
Its form coincides with that of the string (\ref{rate2}) as expected
in a consistent theory. The rate saturates the quantum bound
(\ref{rate4}). But at the time $t_{spread}$ the total length of
string is $L=A/l_P$ and the number of channels becomes equal to the
number of cells with area $\sim l_P^{2}$ on the horizon,
$A/l^{2}_P$, up to a coefficient of the order of unity. This agrees
with a result of Pendry. Pendry showed \cite{pen} that for a system
enclosed within a surface of finite area $A$ the number of channels
is proportional to $A$. On the other hand, our result implies a
limit to the number of channels that can fit into the channel cross
section $A$. It agrees with a result of Lloyd, Giovannetti, and
Maccone \cite{LGM}, who proceeding from a quite different problem
found that the maximum number of channels must be less than
$A/4l^{2}_P$ (our result implies a bound which is four times
larger). Note also that the rate (\ref{rate5}) can be viewed as
follows. By definition, the entropy rate is $S/\tau$, where $\tau$
is the characteristic time of a system. Then, since for a black hole
$\tau \sim R_g$, we obtain at once $dS_{bh}/dt=S_{bh}/R_g$. But this
is nothing else than Lloyd's limit \cite{llo}. According to Lloyd,
this is the maximum rate at which the entropy can be moved in and
out of a system with size $R_g$ and entropy $S$ (attained by taking
all the entropy $S$ in the system and moving it outward at the speed
of light). This has something in common with the observation that
the spreading rate (\ref{exp}) is exactly the fastest rate
consistent with causality \cite{s3}. Thus black holes are the most
extreme objects in nature realizing the maximum possible entropy
flow allowed by quantum theory and relativity.

\emph{Conclusions}.-In this paper we modeled a string spreading
exponentially over the stretched horizon by a self-avoiding walk. In
the framework of this model we found the string entropy and rate at
which it spreads, $dS_s/dt= 2\pi S_s T_H$. We applied two different
quantum limits (Bekenstein-Bremermann and Pendry's) to the rate
leading to the same conclusion: this is the maximum rate allowed by
quantum theory.

\end{document}